\documentclass[twocolumn,epsf,prd,floatfix,nofootinbib]{revtex4}
\usepackage{graphicx}
\usepackage{epsfig}
\usepackage{dcolumn}
\usepackage{bm}
\usepackage{latexsym}

\begin{document}

\title{Tau energy losses at ultra-high energy: continuous versus stochastic treatment.}
\author{Oscar Blanch Bigas$^1$, Olivier Deligny$^2$, K\'evin Payet$^3$, V\'eronique Van Elewyck$^{2,4}$}

\affiliation{
$^1$ LPNHE, Universit\'e Paris VI-VII \& CNRS/IN2P3, Paris, France\\
$^2$ IPN, Universit\'e Paris Sud \& CNRS/IN2P3, Orsay, France\\
$^3$ LPSC, Universit\'e Joseph Fourier Grenoble 1, CNRS/IN2P3, INPG, Grenoble, France\\ 
$^4$ Astroparticule et Cosmologie (UMR 7165) \& Universit\'e Paris 7, Paris, France}

\begin{abstract}
We study the energy losses of the tau lepton in matter through
electromagnetic processes at ultra-high energy (UHE). We use both a
stochastic and a continuous framework to treat these interactions
and compare the flux of tau leptons propagated after some
amount of matter. We discuss the accuracy of the approximation of 
continuous energy losses by studying the propagation in standard rock
of taus with both mono-energetic and power law injection spectra.
\end{abstract}

\maketitle

\section{Introduction}

The observation of high-energy cosmic neutrinos produced in distant
astrophysical sites (or possibly by other, more exotic, mechanisms)
has become one of the major challenges of astroparticle physics. In
both astrophysical and exotic models, substantial fluxes of electron
and muon neutrinos are expected from the desintegration of charged pions
(and kaons) produced in the interaction of accelerated particles with
ambient matter and radiation, either at the source location or on their
way through the universe. Given the large distances traveled by these
cosmic neutrinos, approximately equal fluxes in $\nu_e , \nu_\mu$
and $\nu_\tau$ are expected on Earth as a result of flavour
mixing and oscillations~\cite{nuoscil,athar,halzen}.

At ultra-high energies, the Earth becomes opaque to neutrinos as a 
result of charged- and neutral-current interactions which deplete their 
energy or convert them into the associated charged lepton in case of 
charged-current. Tau neutrinos can then be traced through the observation 
of byproducts of the decay of the associated tau lepton in matter such as 
ice~\cite{ice3,rice,anita}, water~\cite{baikal,km3}, or in the 
air~\cite{auger,hires}.

The tau lepton is subject to energy-dependent radiative processes : 
Bremsstrahlung, production of $e^+ e^-$ pairs and photonuclear 
interactions (ionisation is negligible at ultra-high energy).
A comprehensive knowledge of the probability that a tau with 
initial energy $E_0$ will not be absorbed during its propagation 
in matter is thus required to predict the expected rates of tau 
neutrinos detection by various experiments. In the muon case, 
because of its large lifetime, the survival probability is 
entirely determined by the energy losses over the whole energy range. 
Neglecting the fluctuations in energy losses (or, in other words, 
adopting the continuous energy loss approach) is known to lead to an
overestimate of the survival probability~\cite{lipari}. In contrast,
due to the short lifetime of the tau, the calculation of the range
in that case is determined by the decay up to $\approx10^{8}$ GeV.
Then, the attenuation length becomes smaller than the decay length.
Fluctuations in the energy losses may then play an important
role - as in the muon case.

In this respect, it has already been put forward that such
effects may indeed not be negligible for the tau~\cite{dutta2}.
The approach adopted in this previous work was to inject tau
neutrinos in matter as a source of tau leptons, and then to
compute the flux of tau leptons after some fixed distance of matter.
This calculation was performed in two ways: the first by using
a full simulation of all involved processes through a Monte-Carlo
generator to account for the stochasticity of the problem, and the
second by using a simplified semi-analytical framework to reproduce
the solution within the continuous energy loss framework. Eventual
differences in the flux calculations were attributed to the lack
of accuracy of the continuous energy loss framework.

Such a strategy however makes it difficult to disentangle the direct 
study of $\tau$ propagation from other effects related with the 
propagation and interactions of the $\nu_\tau$. The aim of this paper 
is thus to re-examine the effects of the fluctuations in energy losses 
due to radiative processes of the tau lepton at ultra-high energy in 
a framework where we choose to directly inject $\tau$'s instead of 
$\nu_\tau$'s. This method allows to simplify the problem and to 
get a more direct comparison between the continuous and the
stochastic treatment of the energy losses.

In section \ref{sec:prop}, we present the physical processes that we 
take into account in this study, and the methods we will use to compute 
the propagated flux within both the continuous energy loss framework
and the stochastic one. In section \ref{sec:mono}, we apply our
calculations to the case of mono-energetic beams of tau leptons,
whereas in section \ref{sec:power} we do the same in case of a
power law injection spectra of tau and compare the results obtained
within the two frameworks. Finally, we discuss our conclusions in
section \ref{sec:concl}.

\section{Propagation of tau through matter}
\label{sec:prop}

\subsection{The processes}
\label{subsec:procs}
The tau lepton is an unstable particle. In the observer frame, its 
decay length is energy-dependent :
\begin{eqnarray}
  \lambda_{dec}(E) = 49 \bigg(\frac{E}{1\mathrm{EeV}}\bigg) \mathrm{km}.
\end{eqnarray}

At the energies of interest for this study, the relevant electromagnetic 
processes that the tau lepton undergoes are the Bremsstrahlung, the pair 
production and the photonuclear interactions. Whereas the cross sections 
of the two first ones are well established~\cite{brem,pp}, the 
description of the photonuclear process relies on the proper modelisation
of the nucleon structure functions at low $x$ and high $Q^2$, and several 
calculations of the corresponding cross section can be found in the 
literature. The differences reside in the parameterisation of the
relevant structure functions obtained from different formalisms.
A popular treatment uses a combination of the generalized vector
dominance model for the non-perturbative regime~\cite{BB}, and of
the color dipole model for the perturbative part~\cite{BS}.
Other widely used results are obtained on basis of parameterisations
of data using Regge theory~\cite{dutta1,regge}. A recent comparative study
shows that all these calculations give rates of photonuclear energy loss
in good agreement~\cite{parente}, except for~\cite{PT} which
results in a significantly higher rate at energies above $10^8$ GeV.
A new calculation based on saturation physics~\cite{armesto}
is also presented in~\cite{parente}. In this case the
corresponding energy loss rate is more than a factor of 2
lower than the standard results at the highest ($\sim 10^{12}$ GeV)
energies. In the following, we use the parameterisation
obtained by Dutta et al. \cite{dutta1} for all the calculations
we are presenting.

Finally, the tau lepton is also subject to weak interactions. However, 
using the cross sections derived in~\cite{sarkar}, the corresponding 
interaction length is expected to be much larger than for the 
electromagnetic processes until an energy greater than  $10^{12}$ 
GeV. As we are only interested in studying the effect of the fluctuations 
of the electromagnetic interactions of the tau, we can thus comfortably 
neglect its weak iteractions. We do as well regarding the regeneration 
of $\tau$ leptons at lower energy through the chain 
$\tau \rightarrow \nu_\tau \rightarrow \tau$ via a tau decay
followed by a charged-current weak interaction of the neutrino.

\subsection{Solving the equation of transport}

Within the above assumptions, the transport equation that describes the 
evolution of the $\tau$ flux $\Phi$ along its path trough matter, accounting 
for all processes of creation or absorption of a tau with an energy $E$ at 
a position $x$, is given by:
\begin{eqnarray}
  &&\frac{\partial \Phi(E,x)}{\partial x} =
  -\frac{\Phi(E,x)}{\lambda_{dec}(E)}
  -\sum_i \bigg[
  \frac{\Phi(E,x)}{\lambda_i(E)}
  \nonumber\\
  &-&\frac{\rho \mathcal{N}}{A}
  \int\frac{\mathrm{d}y}{1-y}\Phi
  \bigg(\frac{E}{1-y},x\bigg)
  \frac{\mathrm{d}\sigma_{i}}
  {\mathrm{d}y}\bigg(y,\frac{E}{1-y}\bigg)
  \bigg],
  \label{eq:PhiTau1}
\end{eqnarray}
where $\lambda_{dec}(E)$ is the decay length, $\lambda_i(E)$ and 
$\mathrm{d}\sigma_{i}/\mathrm{d}y(E,y)$ respectively the interaction
length and the differential cross section of each process,  
$y$ the inelasticity of the process considered, $\rho$ the density of 
matter, $\mathcal{N}$ the Avogadro number and $A$ the atomic mass
number.

\subsubsection{Continuous energy losses approximation}

\begin{figure}[t]
    \centering
    \includegraphics[width=9cm,height=9cm]{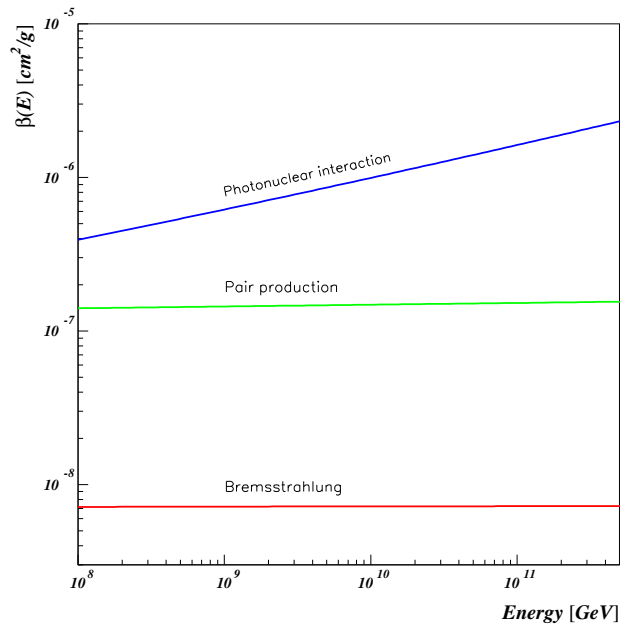}
    \caption{\small{Evolution of $\beta$ as a function of the energy
    for a $\tau$ lepton in standard rock ($A$=22). From bottom to top: 
    bremsstrahlung, pair production and photonuclear
    interactions.}}
    \label{fig:beta}
\end{figure}

When the differential cross-sections exhibit a peak near
$y$=0, as it is indeed the case for the processes we are dealing
with, the integrals are dominated by the behavior of the integrands
around 0, in such a way that an expansion of these integrands can
be performed \cite{landau}. At first order in $y$, this yields
 the following equation :
\begin{eqnarray}
  \frac{\partial \Phi(E,x)}{\partial x} =
  &-&\frac{\Phi(E,x)}{\lambda_{dec}(E)}
  +\rho \frac{\partial}{\partial E}
  \bigg(E\beta(E)\Phi(E,x)\bigg)
  \label{eq:PhiTau2}
\end{eqnarray}
where we have introduced the standard notation
\begin{eqnarray}
  \beta(E)=\frac{\mathcal{N}}{A}\sum_i
  \int_{y_{min}^i}^{y_{max}^i}\mathrm{d}y\,y\,
  \frac{\mathrm{d}\sigma_{i}}{\mathrm{d}y}(E,y).
  \label{eq:beta}
\end{eqnarray}
We show on Fig.\ref{fig:beta} the evolution of $\beta$ as a
function of energy for the three processes described in the
previous sub-section. Note that the contribution of the
energy loss due to the photonuclear interactions does not
reach any asymptotic value because of the expected growth of
the photo-nucleon cross section $\sigma_{\gamma N}$.

\begin{figure}[!t]
    \centering
    \includegraphics[width=9cm,height=9cm]{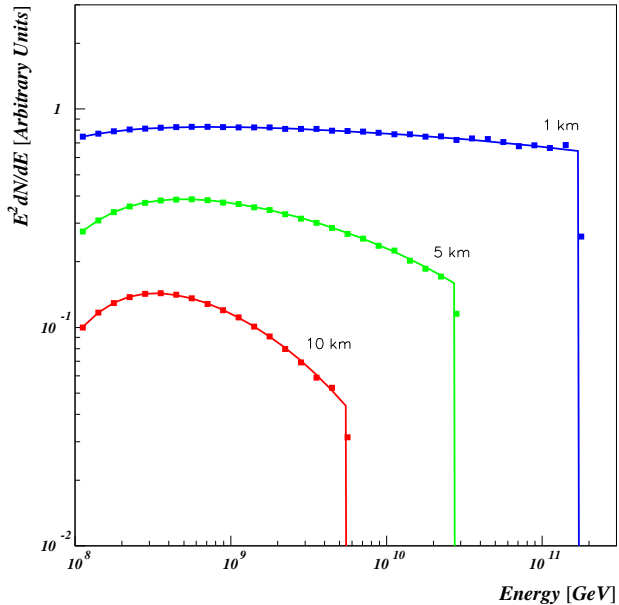}
    \caption{\small{Comparison of the semi-analytical solution
    (solid line) of the transport equation within the continuous
    energy losses approximation (Eqn.~\ref{eq:sol}) with respect
    to a Monte-Carlo calculation (squared points) within the same 
    approximation. An $E^{-2}$ flux
    between $10^8$ GeV and $3\cdot 10^{11}$ GeV is used at injection.
    Three distances of propagation in rock are shown: 1 km,
    5 km and 10 km (from top to bottom).}}
    \label{fig:sa_vs_mc}
\end{figure}

Within the approximation of continuous energy losses,
the average energy lost per unit distance $\mathrm{d}E/\mathrm{d}x$
is, at ultra-high energy, assumed to be proportional to the mean
inelasticity of each process in the following way :
\begin{eqnarray}
	\frac{\mathrm{d}E}{\mathrm{d}x} = -\rho\frac{\mathcal{N}}{A}
	E \sum_i \left<y_i\right> \sigma_i(E).
	\label{eq:dedx}
\end{eqnarray}
The right-hand side of this expression is nothing else but
$-\rho E\beta(E)$. This implies that Eqn.~\ref{eq:PhiTau2} can
be easily integrated, leading to the following expression :
\begin{eqnarray}
	\Phi(E,x) = \Phi_0(\tilde{E}_0)	\exp{\int_0^x
	\mathrm{d}u \bigg(\frac{\partial}{\partial E} \gamma(\tilde{E}_u)
	-\frac{1}{\lambda_{dec}(\tilde{E}_u)} \bigg) }
	\label{eq:sol}
\end{eqnarray}
where $\gamma(E)=\rho E\beta(E)$ and $\tilde{E}_{v}$ the solution of
\begin{eqnarray}
	\int_{\tilde{E}_{v}}^E \frac{\mathrm{d}E_\tau}{\gamma(E_\tau)}=v-x.
\end{eqnarray}
In the following, we use Eqn.~\ref{eq:sol} to compute any propagated flux 
of tau leptons when referring to as the continuous energy losses 
approximation. To verify this expression, we performed a Monte-Carlo 
calculation evaluating at each step the decay probability as well as the 
continuous energy losses through Eqn.~\ref{eq:dedx}. For an incident flux 
of tau leptons following a $E^{-2}$ spectrum in energy between $10^8$ GeV 
and $3\cdot10^{11}$ GeV, we show the agreement of the two calculations on 
Fig.\ref{fig:sa_vs_mc} after a propagation in 1, 5, and 10 km of rock. Let 
us point out that, within the continuous energy losses approximation, the 
sharp cutoffs in energy result from the choice of the maximal energy
at $x=0$, which is univocally related to the propagated energy at depth $x$.

\subsubsection{Stochastic treatment}

To test the accuracy of results obtained by using the
continuous energy losses approximation, we need to account
for the stochastic nature of the transport equation.
A convenient way to solve Eqn.~\ref{eq:PhiTau1}
is to use a Monte-Carlo generator sampling all the interactions.
However, to limit CPU time spent where the cross sections are
large but the energy losses are small, a standard way to proceed
is to separate the losses into two components~\cite{music,dutta2}:
a continuous one where the rate of the losses is large
($y\in[y_{min},y_{cut}]$), and a stochastic one where the
differential cross sections lead to more catastrophic losses
but with a weaker rate ($y\in[y_{cut},y_{max}]$). In practice,
this means that within an elementary step, the Monte-Carlo
samples all the interactions according to cross sections computed
for $y\geq y_{cut}$ together with the probability to decay. At the 
same time, a continuous energy loss is applied according
to Eqn.~\ref{eq:beta}, except that the upper bound of the $\beta$
coefficient is replaced by the cut $y_{cut}$. In that way, the $y$
range where the stochastic nature of the interactions can be
relevant is taken into account, and at the same time, the frequent
small losses due to the peaking of the cross sections near $y=0$
are applied. A good compromise to reproduce the stochastic features
using a reasonably fast code is to take $y_{cut}=10^{-3}$.

\begin{figure}[!t]
    \centering
    \includegraphics[width=9cm,height=9cm]{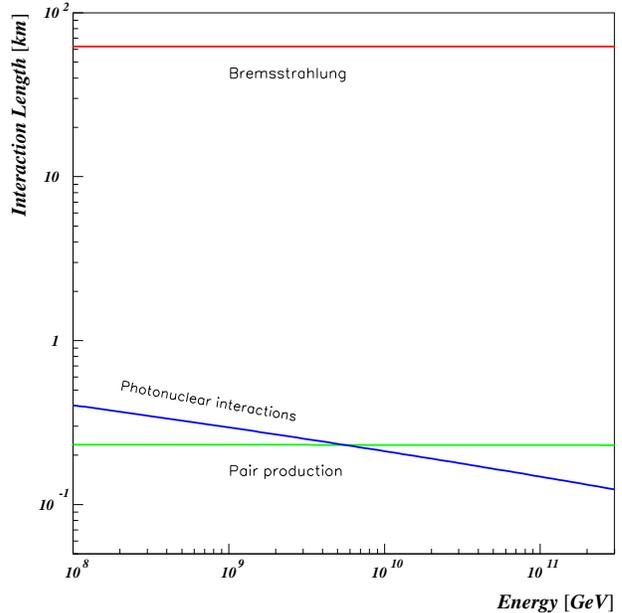}
    \caption{\small{Interaction lengths computed in the $y$ range
    $y\in[y_{cut},y_{max}]$ with $y_{cut}=10^{-3}$.}}
    \label{fig:lint}
\end{figure}

To get a feeling of the effects which may be induced by fluctuations
in energy losses, we plot on Fig.\ref{fig:lint} the interaction length 
of each process, but restricting the $y$ range to $y\in[y_{cut},y_{max}]$. 
The interaction length of the Bremsstrahlung is clearly the higher, so that 
the effects induced by this process are  marginal. On the other hand,
the interaction lengths induced by pair production and photonuclear
interactions are lower and rather close to each other, leading to higher 
rates. Stochastic effects  are not expected to be strong for pair production 
processes, as the corresponding differential cross-section behaves as 
$\approx y^{-2}$. On the contrary, the rate of  photonuclear interactions 
is never negligible, and in particular, above $\approx$ 6 EeV, this 
process becomes the dominant one. Its differential cross section behaves
as $y^{-1+\alpha}$ with $\alpha$ slighty decreasing with energy
from 0.1 at $10^8$ GeV to 0.05 at $10^{12}$ GeV. Hence, fluctuations
in energy losses may become important due to this interaction.

\section{Mono-energetic tau}
\label{sec:mono}

\subsection{Propagated energy spectra}

\begin{figure}[!t]
    \centering
    \includegraphics[width=9cm,height=9cm]{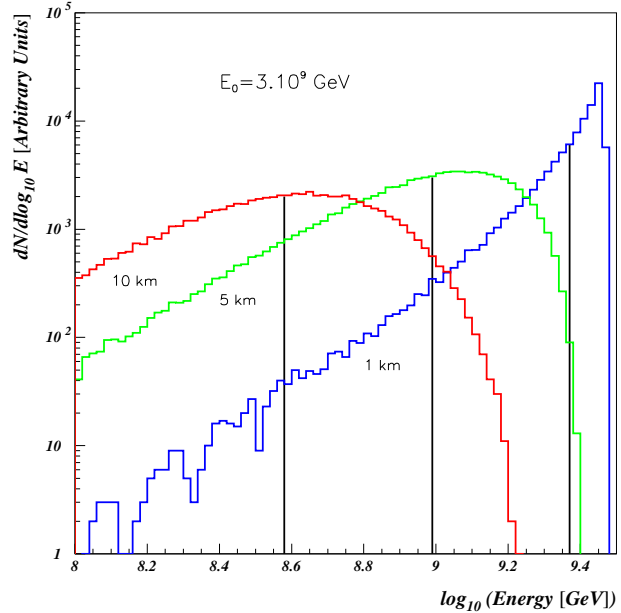}
    \includegraphics[width=9cm,height=9cm]{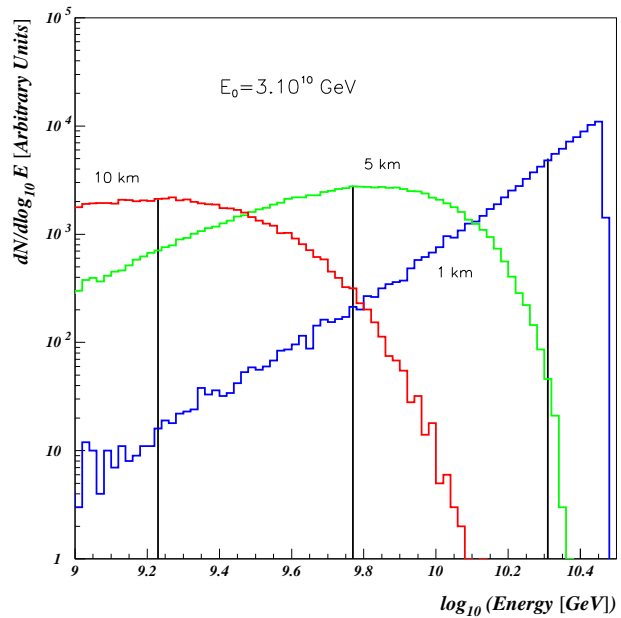}
    \caption{\small{Energy log$_{10}$-distributions of taus of
    $E_0=3\cdot 10^9$ GeV after propagation in standard rock for
    3 different depths (1, 5 and 10 km). Also shown are the mean
    values of the energy distributions.}}
    \label{fig:mono}
\end{figure}

Let us first consider the case of a mono-energetic beam of tau
leptons propagating in rock. The simulated log$_{10}$-distributions 
of the energy of an incident $3\cdot 10^9$ GeV tau beam after crossing 1,
5, and 10 km of standard rock are displayed in Fig.~\ref{fig:mono} (top).
We show on the same plots the means of the energy distributions, which 
are the values used within the continuous energy losses approximation. 
The most probable values are always greater than the mean values of the 
distributions. After 1 km, a good part of the simulated events still 
carry a large fraction of the initial energy $E_0$, meaning 
that this fraction of particles did not undergo many interactions.
However, the distribution is asymmetric and there
is already a long tail of events undergoing hard losses.
For larger paths in rock, fluctuations in the energy losses increase, 
resulting in a broadening of the distribution and a smoothening of 
the high-energy cutoff.

The bottom panel of Fig.~\ref{fig:mono} shows the same quantities for a 
higher initial energy of the incident $\tau$ beam ($E_0 = 3\cdot10^{10}$ GeV). 
As expected from the discussion about the effective interaction lengths for
$y \geq y_{cut}$, those distributions reflect the fact that the
fluctuations are larger at higher energies due to the higher rate of 
photonuclear interactions, which are precisely those leading to harder 
losses.

\subsection{Tau-lepton range}

\begin{figure}[!t]
    \centering
    \includegraphics[width=9cm,height=9cm]{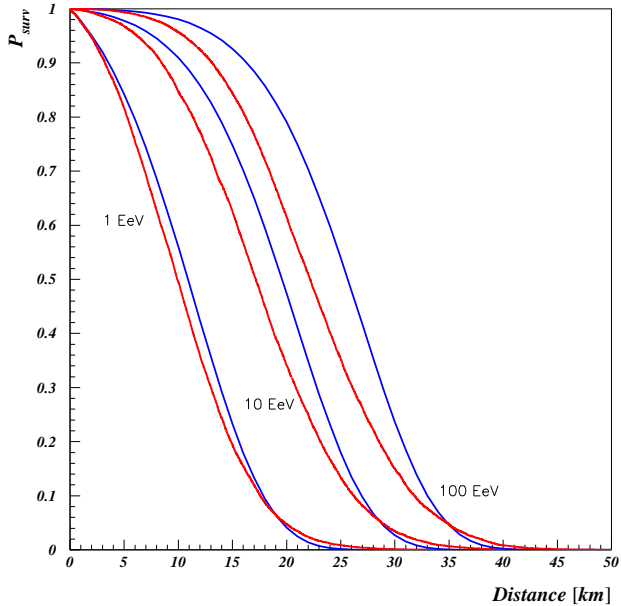}
    \caption{\small{Survival probability in standard rock as a function
    of depth for different initial energies. Blue: continuous energy losses;
    Red: stochastic energy losses.}}
    \label{fig:fluc_range}
\end{figure}

\begin{figure}[!t]
    \centering
    \includegraphics[width=9cm,height=9cm]{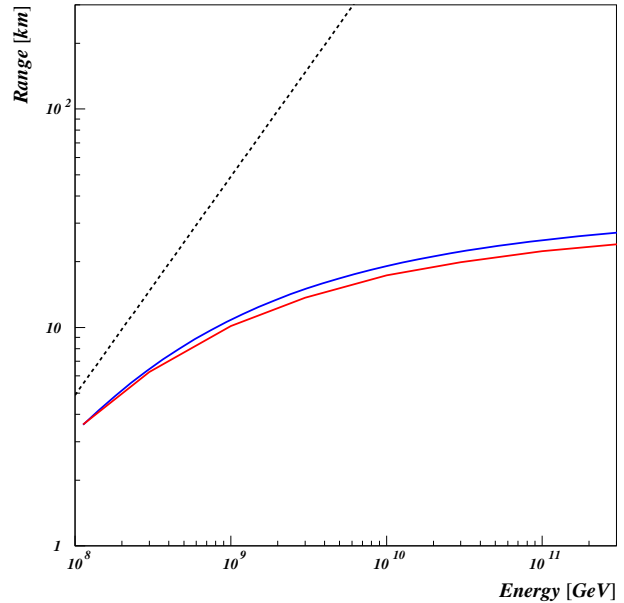}
    \caption{\small{Range $R(E_0)$ of the tau calculated with continuous (top curve) and with stochastic (bottom curve) energy losses. For comparison, the decay length of the tau (dashed line) is also shown.}}
    \label{fig:range}
\end{figure}

For a fixed initial energy of the tau, it is interesting to
calculate the survival probability of the tau as a function
of the traversed depth. In case of continuous energy loss,
the survival probability is simply given by
\begin{eqnarray}
	P_{surv}(E_0,x) = \exp{\bigg(-\int_0^x
	\frac{\mathrm{d}u}{\lambda_{dec}(\overline{E}_u)} \bigg)}
\end{eqnarray}
with $\overline{E}_u$ the solution of
\begin{eqnarray}
	\int_{E_0}^{\overline{E}_{u}} \frac{\mathrm{d}E_\tau}{\gamma(E_\tau)}=-u.	
\end{eqnarray}
We also computed $P_{surv}$ for the case of stochastic losses. We plot in
Fig.~\ref{fig:fluc_range} the resulting curves, showing that in addition
to broadening the distributions, the effect of the fluctuations is
globally to reduce the survival probabilities with increasing energy. 
Note also the presence of a small tail corresponding at higher ranges,
which reflects the cases for which the number of interactions is lower.
This decrease can be quantified through the 
range $R(E_0)$ of the tau lepton which is simply :
\begin{eqnarray}
	R(E_0)=\int_0^\infty \mathrm{d}x \ P_{surv}(E_0,x)
\end{eqnarray}
We show in Fig.~\ref{fig:range} the corresponding range of the tau
lepton in standard rock. Let us remind here that the weak interactions 
of the tau, which we have neglected thorough the whole paper, might change
a little bit the picture at the highest energies. However, rather than giving 
numbers to be taken at face values, we are interested here in analyzing
the effect of the fluctuations, which appear to slightly reduce the range 
of the tau as its energy increases.


\section{Power law injection spectra}
\label{sec:power}

In this section, we consider the case of $E^{-1}$ and $E^{-2}$
tau injection spectra. We restrict ourselves to these generic
fluxes to study possible distorsions on the propagated fluxes
due to the stochastic effects. All the calculations we present
here assume a continuous injection spectrum following a power law
between $10^8$ GeV and $3\cdot 10^{11}$ GeV. The cutoff for the
maximum energy is chosen to be sharp to exhibit most clearly
the different behaviors of the propagated fluxes.

\begin{figure}[t]
    \centering
    \includegraphics[width=9cm,height=9cm]{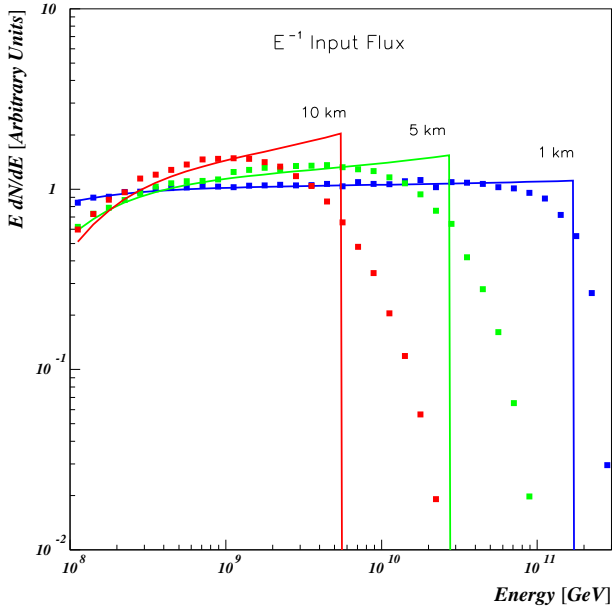}
    \caption{\small{Comparison of the semi-analytical solution
    (solid line) of the transport equation within the continuous
    energy losses approximation (Eqn.\ref{eq:sol}) with respect
    to a Monte-Carlo calculation (squared points) taking into account
    the stochastic effects of the radiative processes. An $E^{-1}$ flux
    between $10^8$ GeV and $3\cdot 10^{11}$ GeV is used at injection.
    Three distances of propagation in rock are shown: 1 km,
    5 km and 10 km (from right to left).}}
    \label{fig:e1}
\end{figure}

\begin{figure}[t]
    \centering
    \includegraphics[width=9cm,height=9cm]{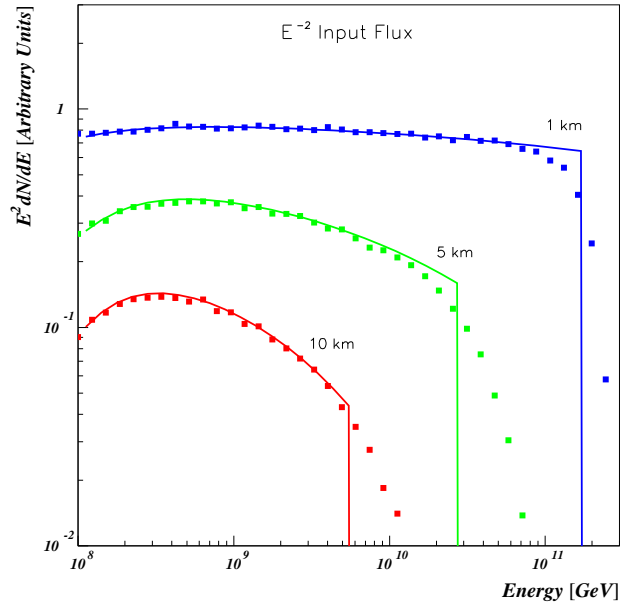}
    \caption{\small{Comparison of the semi-analytical solution
    (solid line) of the transport equation within the continuous
    energy losses approximation (Eqn.\ref{eq:sol}) with respect
    to a Monte-Carlo calculation (squared points) taking into account
    the stochastic effects of the radiative processes. An $E^{-2}$ flux
    between $10^8$ GeV and $3\cdot 10^{11}$ GeV is used at injection.
    Three distances of propagation in rock are shown: 1 km,
    5 km and 10 km (from right to left).}}
    \label{fig:e2}
\end{figure}

In Figs.~\ref{fig:e1} and ~\ref{fig:e2}, we show the results obtained 
respectively from  $E^{-1}$ and  $E^{-2}$ injection spectra. For a better 
lecture of the different structures, we choose to plot the fluxes 
corrected for the injection spectrum. This means that in a trivial 
situation with no losses and no decay of the tau, all results would be 1. 
Continuous lines are from the semi-analytical solution in the approximation 
of continuous energy losses, whereas squared dots are the results of the 
Monte-Carlo simulation including stochastic effects. Again, the sharp 
cutoff present in the semi-analytical solutions results from the choice 
of the maximal energy together with the univocal relation between this 
maximal energy at $x=0$ and the propagated energy at a depth $x$. On the 
other hand, the stochastic treatment of the tau propagation has a smoothening 
effect on this cutoff: there are indeed fluctuations affecting a small fraction 
of particles which undergo less interactions and less hard losses. The energy 
range on which this broadening occurs clearly increases with the depth 
traversed by the tau, and it is also more pronounced in the case of an 
$E^{-1}$ flux. At the same time, we can see from both figures that this 
effect is the only important distorsion induced by the stochastic processes. 
This means that, as far we are dealing with continuous incident spectra 
behaving as power laws, there is a compensation between positive and negative 
fluctuations of the energy losses everywhere in the considered energy range, 
except near the high-energy border. This compensation allows to use the 
continuous energy loss approximation as the correct mean value of the 
propagated spectrum.

\section{Conclusions}
\label{sec:concl}

In this paper, we have shown that the stochastic nature of the radiative 
processes undergone by tau leptons at ultra-high energy is indeed responsible 
for large fluctuations in the tau energy losses. At the same time, however, 
these fluctuations are not so large as to blur the picture with respect to 
the continuous energy loss approximation, as far as the calculation concerns 
power-law injection spectra in a given, continuous energy range.

As already pointed out in Sec.~\ref{subsec:procs}, significant theoretical 
uncertainties exist in the calculations of the cross section for photonuclear 
interactions, which are the most relevant processes for tau energy losses at 
the ultra-high energies we are interested in. The present study was done on 
basis of one particular model~\cite{dutta1} which is rather pessimistic in 
the $y$ range where the stochastic effects can lead to hard losses, in the 
sense that it leads to a fairly high rate of interactions. Any other model 
leading to a lower rate would not change the picture. Moreover, any other 
model leading to comparable or slightly greater rate would have to present 
a significantly harder differential cross section in term of $y$ to challenge 
our conclusions.

\section*{Acknowledgments}
O.B.B. is supported by the Ministerio de Educacion y Cienca of Spain through
the postdoctoral grant program. V.V.E. acknowledges support from the European 
Community 6$^{th}$ Framework program through the Marie Curie Fellowship 
MEIF-CT-2005 025057.

\end{document}